\documentclass[aps, prd, twocolumn,superscriptaddress,longbibliography,numerical]{revtex4-1}

\usepackage{dsfont}
\usepackage{amsmath}
\usepackage{epsfig}
\usepackage{graphicx}
\usepackage{bm}
\usepackage{amssymb}
\usepackage{slashed}
\usepackage[colorlinks,citecolor=blue,linkcolor=blue,urlcolor=blue,plainpages=false,pdfpagelabels]{hyperref}

\begin{document}
% ==============================================================================

\title{Superconductivity from incoherent Cooper pairs in the strong-coupling regime} 

\author{A.~A.~Zyuzin}
\affiliation{---}

\author{A.~Yu.~Zyuzin}
\affiliation{Ioffe Physical--Technical Institute,~194021 St.~Petersburg, Russia}

% ==============================================================================
\begin{abstract}
We propose a scenario for superconductivity at strong electron-electron attractive interaction, in the case when
the increase of the interaction strength promotes the nucleation of the local Cooper pairs and forms a state with a spatially phase incoherent Cooper pair order parameter. 
We show that this state can be characterized by a pseudogap which is determined by the electron scattering by phase fluctuations.
At low temperatures, however, long-range correlations between the regions with different phases become important and establish global phase coherence and hence superconductivity in the system. 
We develop a mean-field theory to describe a phase transition between the preformed Cooper pair and superconducting states. 
This scenario of superconductivity applies not only to conductors with parabolic bands but also to the flat-band systems in which flat 
and dispersive bands coexist and are responsible for the formation of Cooper pairs as well as their 
phase synchronization.
\end{abstract}
\maketitle

% ==============
 \section{\label{sec:intro}Introduction}
% ==============
The study of superconductivity in systems with strong electron-phonon \cite{Migdal, Eliashberg} or non-retarded pairing \cite{JLTP_Schmitt} interactions encounters the problem of ultraviolet divergence, which is commonly addressed by introducing a regularized electron-electron scattering length \cite{Gorkov_Tc, Barkhudarov_Tc}.
The regularized theory \cite{Engelbrecht} is used to discuss the crossover between weak Bardeen-Cooper-Schrieffer (BCS) and strong  Bose-Einstein condensation (BEC) regimes, for instance, in ultra cold atoms and nuclear systems, as reviewed in \cite{Ohashi_progress, Pitaevskii_RevModPhys, Strinati_physrep}. However, strictly speaking, this theory is justified only in cases where the scattering length of the pairing interaction potential is smaller than the Fermi wave length \cite{JLTP_Schmitt}, or when the Fermi energy exceeds the Debye energy, as in the conventional BCS model. In this paper, we revisit the problem of ultraviolet divergence in superconductors.

We investigate a scenario of superconductivity with strong attractive interaction between electrons in the intermediate BCS to BEC crossover regime.
Specifically, we consider a situation where the normal metal state becomes unstable towards the formation of local Cooper pairs
at temperatures much higher than the superconducting transition temperature.  

In the following, we argue that the ultraviolet divergence describes the physics of short-range correlations, indicative of a preformed Cooper pair state.
This state can be understood as a system of small domains, each on the scale of the Fermi wavelength, where the pairing order parameter exhibits spatially uncorrelated random phases.
The coupling of these domains via Andreev reflections leads to phase synchronization and superconductivity at elevated temperatures  \cite{Zyuzin}. 
Our model bears a resemblance to the situation in which Cooper pairs preform due to spatial fluctuations induced by disorder \cite{Nature_Sacepe_Feigelman}, and
also aligns with the case of flat-band superconductivity, \cite{Zyuzin_Zyuzin, Zyuzin_Firoz}. 

The study of flat-band materials is considered to have significant importance for understanding the mechanisms of high-temperature superconductivity \cite{Volovik_review}. One of the key reasons for this is the singular density of states at the flat band, which may support elevated superconducting transition temperatures \cite{Khodel_Shaginyan, Volovik_1994, PhysRevLett_Masanori, Physica_C_Furukawa, PhysRevB_Kopnin_Heikkila_Volovik, Peotta_Torma}. 

Recently, it was recognized that the flatness of the band dispersion guarantees physics similar to that of the strong-interaction regime of Cooper instability \cite{Zyuzin_Zyuzin}. In particular, the ultraviolet cutoff for the divergent superconducting condensation energy can naturally be determined by the width of the flat-band in momentum space \cite{PhysRevB_Kopnin_Heikkila_Volovik}, which, in the extreme limit, extends up to the edges of the first Brillouin zone \cite{Peotta_Torma}. Thus, in addition to an elevated superconducting transition temperature, a non-BCS type of phase transition in flat-band systems may be expected.

Furthermore, it was recently noted that the spatial locality of flat-band electrons can serve as a precursor to a pseudogap state, characterized by an uncorrelated random phase of the Cooper pair order parameter \cite{Zyuzin_Zyuzin, Zyuzin_Firoz}. In the flat-band scenario, superconductivity emerges due to long-range coupling between preformed pairs, ensured by the dispersive bands, which inevitably correlate with the flat bands in realistic materials \cite{Zyuzin_Zyuzin, Zyuzin_Firoz}. 

Here, in the case of conductors with parabolic bands and strong attractive interaction, we find that the pairing instability is governed by two physically distinct length scales of electron-electron correlations: short-range correlations, which support local pair binding at high temperatures, and long-range correlations, which drive phase synchronization between these preformed pairs at lower temperatures.

The rest of this paper is organized as follows: In Section (\ref{sec:model}), we introduce a model for superconductivity in the strong interaction limit. We emphasize the significance of short- and long-range correlations in the paring instability and discuss solutions to the self-consistency equation for the order parameter in both weak- and strong- interaction regimes. In Section  (\ref{sec:mean_field}), we present the mean-field theory of superconductivity in the strong coupling case. In Section (\ref{sec:pseudogap}), we provide explicit calculations for the electron self-energy, due to scattering by random phase fluctuations. Finally, our findings are summarized in Section (\ref{sec:conclusion}).

% ==============
 \section{\label{sec:model}Model}
% ==============
We consider a three-dimensional electron gas system in the presence of an attractive interaction between electrons. The conduction-band electrons
with parabolic dispersion are described by the Hamiltonian 
$
H =  \int d\mathbf{r}\psi_{\sigma}^\dag(\mathbf{r}) [-\boldsymbol{\nabla}^2_r/(2m) -\mu] \psi_{\sigma}(\mathbf{r}),
$ 
where $\mu>0$ and $m$ are the chemical potential and effective mass of electrons, $\psi^{\dag}_{\sigma}(\mathbf{r}), \psi_{\sigma}(\mathbf{r})$ are the electron operators with spin projection $\sigma = \uparrow, \downarrow$ ($\hbar=k_{\mathrm{B}}=1$ units are used henceforth).  The interaction Hamiltonian is given by
\begin{eqnarray}\label{Model_potential}
H_{\mathrm{int}} =  \int d\mathbf{r}d\mathbf{r}'\psi_{\uparrow}^\dag(\mathbf{r}) \psi_{\downarrow}^\dag(\mathbf{r}') U(\mathbf{r}-\mathbf{r}') \psi_{\downarrow}(\mathbf{r}')  \psi_{\uparrow}(\mathbf{r}),~~
\end{eqnarray} 
where the interaction is modelled as a potential well of size $a$ and depth $U>0$,
\begin{equation}\label{Model_potential}
U(\mathbf{r}) = - U \theta(a-r).
\end{equation} 
We consider a short-range potential with a radius smaller than the Fermi wavelength of electrons, $a<\lambda_{\mathrm{F}}\equiv2\pi/\sqrt{2m\mu}$. 
In the BCS theory, conventionally, the pairing potential is assumed to be local. However, in the strong-coupling limit, it is necessary to incorporate finite-size effects to accurately account for ultraviolet divergence. 

As derived in the Appendix (\ref{sec:Appendix_A}), the potential (\ref{Model_potential}) creates a two-particle bound state, provided that
\begin{equation}\label{main_statement}
U> \frac{\pi^2}{4 m a^2}\equiv U_c =\frac{\mu}{2}\left(\frac{\lambda_{\mathrm{F}}}{2a}\right)^2\geq \mu.
\end{equation}
This means that the system becomes unstable towards the local pair binding at the critical strength of attraction $U=U_c$. 

Interestingly, the increase in the mass of quasiparticles suppresses the interaction threshold. Thus, the critical interaction strength must be smaller in heavy-fermion materials or artificial flat-band systems. 

Mathematically, the emergence of pair instability can be uncovered by investigating the divergent poles in the particle-particle scattering ladder or equivalently by demonstrating the existence of a nontrivial solution for the order parameter $\Delta(\mathbf{R},\mathbf{r}) $ in the equation (see the Appendix (\ref{sec:Appendix_A}))
\begin{equation}\label{Main_equation}
\Delta(\mathbf{R},\mathbf{r})= U \int d\mathbf{R}' \int_{r'\leq a } d\mathbf{r}' \Pi\left(\mathbf{R}-\mathbf{R}',\frac{\mathbf{r}-\mathbf{r}'}{2}\right)\Delta(\mathbf{R}',\mathbf{r}'),
\end{equation}
where $\mathbf{R}$ and $\mathbf{r}$ denote the center-of-mass and relative coordinates, respectively. The lengths of lowercase vectors $(\mathbf{r}, \mathbf{r}')$ are bounded by the radius of the potential, $r, r' \leq a$. The kernel of integral equation (\ref{Main_equation}) is defined as
\begin{equation}\label{Pi_def}
\Pi(\mathbf{R},\mathbf{r}) = T\sum_{\omega_n}G_{\omega_n}(\mathbf{R}+\mathbf{r})G_{-\omega_n}(\mathbf{R}-\mathbf{r}),
\end{equation} 
where $T$ is the temperature and the Green function in the Matsubara frequency $\omega_n =(2n+1)\pi T$  and spatial coordinate representation $G_{\omega_n}(\mathbf{R}) = \int \frac{d^3p}{(2\pi)^3} \exp(i\mathbf{p}\cdot\mathbf{R})(i\omega_n- \xi_{\mathbf{p}})^{-1}$, where $\xi_{\mathbf{p}} = \mathbf{p}^2/2m - \mu$, is given by
\begin{eqnarray}\label{Green_function}
G_{\omega_n}(\mathbf{R}) = -\frac{m}{2\pi R} \exp\left\{i\mathrm{sign}(\omega_n)\sqrt{1+\frac{i\omega_n}{\mu}}\frac{2\pi R}{\lambda_{\mathrm{F}}}\right\}.
\end{eqnarray}

% ==============
\subsection{\label{sec:separation}The separation of scales}
% ==============

The calculation of the kernel (\ref{Pi_def}) reveals a separation of scales in the superconductivity problem for $a<\lambda_{F}$, as we shall demonstrate below.
This separation arises from the asymptotic coordinate dependence of $\Pi(\mathbf{R},\mathbf{r}) $ on small and large length scales relative to $\lambda_{F}$.

For $\lambda_{F}> R > r\approx a$, in the Appendix (\ref{sec:Appendix_A}) it is shown that the kernel (\ref{Pi_def}) diverges as
\begin{equation}\label{Loc_kernel}
\Pi(\mathbf{R},\mathbf{r}) =\frac{m}{(2\pi)^3} \frac{1}{(R^2+r^2/4)^2} -\frac{m^{3}\lambda_{F}T^{2}}{24\pi^{2}R}.
\end{equation}
This represents the short-range part of the kernel, which provides the diverging contribution to (\ref{Main_equation}) at small distances. 
The first term in expression (\ref{Loc_kernel}) can be evaluated by setting the chemical potential in the Green functions (\ref{Green_function}) to zero, as in the problem of two-particle scattering. 
The second term describes a temperature-dependent correction. We note that expression (\ref{Loc_kernel}) is cut off at short distances $r\sim a$ by the radius of the pairing potential.

In the BCS weak-coupling limit, the local contribution does not lead to Cooper instability and is conventionally absorbed into the renormalized interaction constant (see, for example, \cite{Gorkov_Tc, Barkhudarov_Tc}). 
 On the other hand,  in the strong interaction limit, local correlations are detrimental to the pairing instability, with flat-band superconductivity being a primary example \cite{Volovik_review}.

For $R > \lambda_{\mathrm{F}}$, the long-range part of the kernel is given by
\begin{equation}\label{Kernel_interm}
\Pi(\mathbf{R},\mathbf{r}) = \frac{m}{2\pi\lambda_{\mathrm{F}}R^2\ell_{\mathrm{T}}} \frac{1}{\sinh(2\pi R/\ell_{\mathrm{T}})}.
\end{equation}
Physically, it is responsible for the long-range correlations between Cooper pairs, as it decays exponentially over the thermal length 
$\ell_{\mathrm{T}}= v_{\mathrm{F}}/T$. Mathematically, at large distances ($R > \lambda_{\mathrm{F}}$), the dependence on $r$ in (\ref{Pi_def}) can be neglected, allowing us to use the notation $\Pi(\mathbf{R})$ instead of $\Pi(\mathbf{R},\mathbf{r})$. Taking this into account, at length scales smaller than the thermal length $\ell_{\mathrm{T}}>R$, the expression in (\ref{Kernel_interm}) can be brought to the form
\begin{equation}\label{Nonloc_kernel}
\Pi(\mathbf{R}) = \frac{m}{4\pi^2\lambda_{\mathrm{F}}R^3}.
\end{equation}

To sum up, we find that the kernel (\ref{Pi_def}) contains two qualitatively distinct asymptotic behaviors, (\ref{Loc_kernel}) and (\ref{Nonloc_kernel}), which describe short- and long-range correlations between particles in the Cooper channel.
We want to highlight that this separation of scales is strikingly comparable to what is observed in flat-band systems, making the latter resemble the case of a simple parabolic electron energy dispersion.
However, in quasi-flat-band systems, the finite energy width of the band may lead to a dome like dependence of various physical observables, such as the critical temperature, on doping, an effect that differs from the present case. 

% ==============
 \subsection{\label{sec:weak_strong}Weak and strong interactions}
% ==============
Let us now investigate the manifestation of scale separation in the weak and strong interaction regimes. 
In the Appendix (\ref{sec:Appendix_A}), we show that solution of (\ref{Main_equation}) can be represented as
\begin{equation}\label{Delta_separation}
\Delta(\mathbf{R},\mathbf{r}) = \frac{ \Delta(\mathbf{R})}{r} \sin\left(\frac{\pi r}{2a}\sqrt{\frac{U}{U_c}}\right),
\end{equation}
where $\Delta(\mathbf{R})$ changes slowly on the scale of the wave-length $\lambda_{\mathrm{F}}$, while the $r$-contribution describes the position dependence inside the potential well as $r<a$.
We will investigate the self-consistency equation for $ \Delta(\mathbf{R})$ in the weak- and strong- interaction regimes by considering the local and nonlocal parts of the kernel as perturbations, respectively.

In the Appendix (\ref{sec:Appendix_A}), from Eq. (\ref{Main_equation}), we derive an equation that governs the superconducting transition temperature at $\mu\gg T$:
\begin{equation}\label{SCE_weak_int}
\Delta(\mathbf{R}) = \frac{U}{U_\mathrm{c}}\left\{1+ \frac{32a}{\pi^2 \lambda_{\mathrm{F}}}\ln\left|\frac{\ell_{\mathrm{T}}}{\pi \lambda_{\mathrm{F}}} \right|\right\}\Delta(\mathbf{R}).
\end{equation}
The first term on the right-hand side of Eq. (\ref{SCE_weak_int}) comes from the local contribution (\ref{Loc_kernel}) after integration over lengths smaller than $\lambda_{\mathrm{F}}$ (temperature-dependent corrections were neglected here). 
The second term represents the BCS logarithm, derived under the condition that integration over the coordinate is bounded by $\ell_{\mathrm{T}}$ and $\lambda_{\mathrm{F}}$ from above and below, respectively.

The solution to  Eq. (\ref{SCE_weak_int}) exists provided $U/U_c < 1$. 
It is due to the logarithm contribution at low temperatures, $\ell_{\mathrm{T}}>\lambda_{\mathrm{F}}$. This is the weak-interaction regime. Thus, the metal to superconductor transition temperature is given by the BCS expression
\begin{equation}\label{Tc_naive}
T_c \simeq \frac{\mu}{\pi^2} \exp\left\{- \frac{1-U/U_c}{ \nu U \Omega_a} \right\}, 
\end{equation}
where $\Omega_a = 4\pi a^3/3$ is the volume of the potential well (\ref{Model_potential}) and $\nu = m p_{\mathrm{F}}/2\pi^2$ is the electron density of states per spin, with $p_{\mathrm{F}}=\sqrt{2m\mu}$ being the Fermi momentum. The local contribution can be incorporated into the renormalized interaction constant, $U/(1-U/U_c)$, making the exponent proportional to the inverse electron-electron scattering length \cite{Gorkov_Tc, Barkhudarov_Tc}.

However, as the electron-electron attraction increases ($U/U_c \rightarrow 1$), it becomes apparent that $\ell_{\mathrm{T}}$ must be on the order of $\lambda_{\mathrm{F}}$ to fulfill Eq. (\ref{SCE_weak_int}). In this case, the correction stemming from the local contribution in Eq. (\ref{SCE_weak_int}) becomes significant, suggesting a reevaluation of Eq. (\ref{Main_equation}). 

\begin{figure}[h]
\centering
\includegraphics[width=6cm]{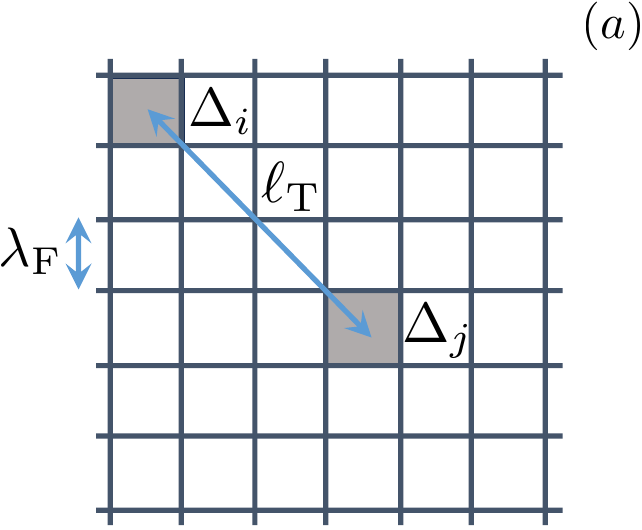}\\
\includegraphics[width=7cm]{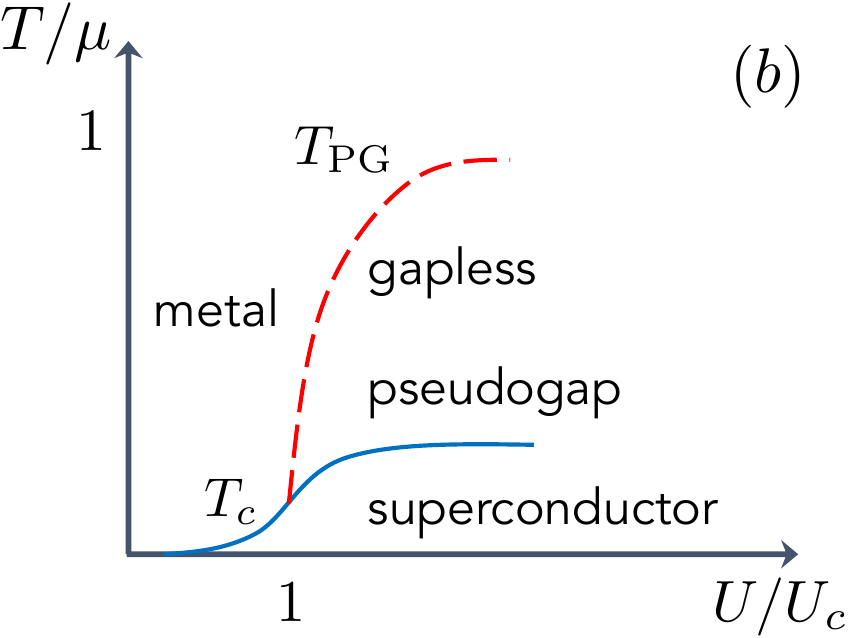}
\caption{\label{fig1} (a) Schematics of the system divided into equal domains with volume $\lambda_{\mathrm{F}}\times \lambda_{\mathrm{F}} \times \lambda_{\mathrm{F}}$, where $\lambda_{\mathrm{F}}$ is the Fermi wave-length. At strong interaction $U>U_c$, one can define the local order parameter $\Delta_i$  in each domain separately. The phases of the order parameter in different domains are uncorrelated. This is the high-temperature pseudogap state. The correlations between the domains on larger length scales $\propto \ell_{\mathrm{T}} = v_{\mathrm{F}}/T$ establish the phase coherence in the system at low temperatures. 
(b)  Schematics of the phase diagram in the plane with a normalized temperature $T/\mu$ and interaction constant $U/U_c$. One distinguishes between the normal metal, superconductor, and phase-incoherent gapless and pseudogap states. The crossover between the normal metal and pseudogap states (dashed curve) is defined by temperature $T_{\mathrm{PG}}$. The superconducting transition is defined by temperature $T_c$ (solid curve). The BCS regime is established in the weak interaction limit, while it changes to a phase-synchronization regime at strong interaction. }
\end{figure}

 In our model, under the strong-coupling conditions, the system partitions into domains with volumes proportional to $\lambda_F^3$, as illustrated in Fig. \ref{fig1}. Within each domain, the order parameter
can be independently identified, subject to a random phase. At $U/U_c \geq 1$, we find that the equation for the order parameter takes the form, see the Appendix (\ref{sec:Appendix_A})
\begin{equation}\label{SCE}
\Delta(\mathbf{R}) = \frac{U}{U_\mathrm{c}}\left\{1-\frac{(4\pi)^{2}aT^{2}}{3\lambda_{\mathrm{F}}\mu^{2}}\right\}\Delta(\mathbf{R}).
\end{equation}
As a result, at $U\gtrsim U_c$, we obtain the crossover temperature for the formation of the local pairing instability
\begin{equation}\label{Tg}
T_{\mathrm{PG}} = \frac{\mu}{4\pi}\left(\frac{3\lambda_{\mathrm{F}}}{ a}\right)^{1/2} \left(1- \frac{U_c}{U}\right)^{1/2}.
\end{equation}
In the derivation of Eq. (\ref{SCE}), we neglect the long-range nonlocal correlations (\ref{Nonloc_kernel}), as they contribute at length scales greater than $\lambda_{\mathrm{F}}$ and average out due to the random phase of the order parameter at such scales. We will later investigate the role of nonlocality in establishing long-range correlations among the domains.

We emphasize that the local nature of the strong-coupling instability implies that the system is not in a phase-coherent superconducting state but, rather, consists of preformed Cooper pairs with spatially uncorrelated random phases.
As the temperature decreases, these correlations establish global phase coherence. 
In the next section, we will analyze the superconducting transition temperature within the framework of the preformed Cooper pair model.

% ==============
\section{\label{sec:mean_field} Mean-field theory}
% ==============
The Hubbard-Stratonovich transformation allows us to reduce the four-fermion interaction, in the stationary approximation, to a functional over the complex bosonic field $\Delta(\mathbf{r}_1;\mathbf{r}_2)$ 
(here, we consider the static approximation with a time-independent $\Delta$):
\begin{align}\label{Fist_action}\nonumber
&S= \int_{\mathbf{r},\tau} 
\overline{\Psi}_{\sigma}(\mathbf{r},\tau)\left[\partial_\tau - \frac{\boldsymbol{\nabla}^2}{2m}-\mu \right]\Psi_{\sigma}(\mathbf{r},\tau)~~\\\nonumber
&+\int_{\mathbf{r}_1,\mathbf{r}_2,\tau}\bigg\{ \frac{|\Delta(\mathbf{r}_1;\mathbf{r}_2)|^2}{U}-\Delta^*(\mathbf{r}_1;\mathbf{r}_2) \Psi_{\downarrow}(\mathbf{r}_2,\tau)\Psi_{\uparrow}(\mathbf{r}_1,\tau)\\
&-\Delta(\mathbf{r}_1;\mathbf{r}_2) \overline{\Psi}_{\uparrow}(\mathbf{r}_2,\tau)\overline{\Psi}_{\downarrow}(\mathbf{r}_1,\tau)\bigg\}\theta(a-|\mathbf{r}_1-\mathbf{r}_2|).
\end{align}
It suffices to separate the spatial variables $(\mathbf{r}_1+\mathbf{r}_2)/2 = \mathbf{R}$ and $\mathbf{r}_1-\mathbf{r}_2 = \mathbf{r}$. After integrating out the fermions $\overline{\Psi}_{\sigma}(\mathbf{r},\tau), \Psi_{\sigma}(\mathbf{r},\tau)$ in (\ref{Fist_action}), we arrive at the bosonic action for
$\Delta(\mathbf{R},\mathbf{r})$. 

In the strong-interaction case due to the local contribution the typical size of the Cooper pair is of the order of the Fermi wavelength.
Hence, we divide the space volume into equal domains with volume $\lambda_{\mathrm{F}}\times \lambda_{\mathrm{F}}\times \lambda_{\mathrm{F}}\equiv \Omega_{\mathrm{F}}$. 
We choose to center these domains at points $\mathbf{R}_N$ and numerate them as $N=1,2,3...$ 
Using  (\ref{Delta_separation}), we discretize the field $\Delta(\mathbf{R})$ by introducing $\Delta(N) \equiv \Delta(\mathbf{R}_N)$ in the $N$-th domain and integrate over the coordinate $\mathbf{r}$ around points $\mathbf{R}_N$ in the bosonic action. 
In this model, $\Delta(N)$ has a random uncorrelated phase in each domain. 
Including the normalization term $\Omega_a$ in the definition of $\Delta(N)$, we obtain the following expression:
\begin{eqnarray}\label{MAIN_action}\nonumber
T S &=& \sum_{N} \bigg\{ \frac{\Omega_{\mathrm{F}}}{\Omega_a}\left(\frac{1}{U} - \frac{1}{\mathcal{U}_c(T)} \right) |\Delta(N)|^2+b  |\Delta(N)|^4
\\
&-&  \Omega^2_{\mathrm{F}}\sum_{N'\neq N}\Delta(N)\Pi(N,N')\Delta^*(N') \bigg\},
\end{eqnarray}
where $\mathcal{U}_c(T) = U_c\left\{1-\frac{(4\pi)^{2}aT^{2}}{3\lambda_{\mathrm{F}}\mu^{2}} \right\}^{-1}$.  The derivation of the quadratic term is given in the Appendix (\ref{sec:Appendix_A}). We also take into account the quartic term $b \propto 1/\mu^3$ as the 
$1/U - 1/\mathcal{U}_c(T)$ term can change sign. The local contribution to the quartic term is given by
\begin{equation}
b=T\sum_{\omega_n} \int_{r_{i}<\lambda_{\mathrm{F}}} G_{\omega_n}(\mathbf{r}_{12})G_{-\omega_n}(\mathbf{r}_{23})G_{\omega_n}(\mathbf{r}_{24})G_{-\omega_n}(\mathbf{r}_{41}),
\end{equation}
where $\mathbf{r}_{ij} = \mathbf{r}_{i}-\mathbf{r}_{j}$ is introduced for brevity and  $\int_{r_{i}<\lambda_{\mathrm{F}}} $ is shorthand notation for the integrals over the coordinates $\mathbf{r}_{i}$ bounded by the Fermi wave-length.

The last term on the right-hand side of (\ref{MAIN_action}) describes the interaction between regions $N$ and $N'$. At $\lambda_{\mathrm{F}}< |\mathbf{R}_N-\mathbf{R}_{N'}|$, the kernel is given by
\begin{equation}\label{Pi_nonlocal}
\Pi(N,N') = \nu\frac{ \mathrm{csch}(2\pi |\mathbf{R}_N-\mathbf{R}_{N'}|/\ell_{\mathrm{T}})}{2\ell_{\mathrm{T}} |\mathbf{R}_N-\mathbf{R}_{N'}|^{2}}.
\end{equation}
Expression (\ref{Pi_nonlocal}) describes the long-range interaction between the domains, which is exponentially suppressed at lengths larger than the thermal length $\ell_{\mathrm{T}}$.

%%%%%%%%%%%%%%%%%%%%%% BEGIN FIGURE %%%%%%%%%%%%%%%%%%%
%%%%%%%%%%%%%%%%%%%%% END FIGURE %%%%%%%%%%%%%%%%%%%

In the vicinity of temperature (\ref{Tg}), using (\ref{MAIN_action}), we find the amplitude of the local order parameter $\Delta(N) = |\Delta_{\mathrm{PG}}| \exp[i\phi(N)]$ in the form
\begin{equation}\label{mean_field_gap_function}
|\Delta_{\mathrm{PG}}|^2= \frac{1}{2b}\frac{ \Omega_{\mathrm{F}}}{\Omega_a}\left(\frac{1}{\mathcal{U}_c(T)}-\frac{1}{U}\right).
\end{equation}
Restoring the temperature dependence, for $T\rightarrow T_{\mathrm{PG}}$ we obtain
\begin{equation}
|\Delta_{\mathrm{PG}}| \propto T_{\mathrm{PG}}\sqrt{1 - T^2/T^2_{\mathrm{PG}}}.
\end{equation}
The limiting case $T\ll \mu$ considered here allows us to truncate the $T/\mu$ expansion in Eq. (\ref{SCE}) at the lowest nontrivial order, leading to the condition on the interaction region: 
$
1- U_c/U \gtrsim a/\lambda_{\mathrm{F}} 
$ in the strong coupling case. As the interaction strength $U$ increases, higher-order terms in $T/\mu$ in the self-consistency equation (\ref{SCE}) become comparable, breaking down our assumption $T/\mu \ll 1$.

Having identified the amplitude of the local order parameter and the crossover temperature for local instability, let us now investigate the transition between a preformed Cooper pair state and the superconducting state tuned by long-range correlations (\ref{Pi_nonlocal}).

At the superconducting phase transition, we have a nonzero average for the order parameter, defined as
\begin{eqnarray}\label{MFSC_equation}
\langle \Delta(N) \rangle = \frac{\int D[\Delta, \Delta^*] \Delta(N) \exp\{-S[\Delta, \Delta^*]\}}{\int D[\Delta, \Delta^*] \exp\{-S[\Delta, \Delta^*]\}}.
\end{eqnarray}
To evaluate the superconducting transition temperature in Eq. (\ref{MFSC_equation}), we develop the mean-field theory of superconductivity, mapping it to the Langevin theory of magnetism \cite{Malshukov, Zyuzin}. 
Namely, substituting $\Delta(N)$ by its mean value $\langle \Delta(N) \rangle$ in each domain except $N=0$, allows us to obtain the mean-field approximation for the action $S[\Delta, \Delta^*]$ as
\begin{align}\label{Mean_field_functional}
&T S_{\mathrm{MF}} = \left(\frac{1}{U} - \frac{1}{\mathcal{U}_c(T)} \right)\frac{\Omega_{\mathrm{F}}}{\Omega_a} |\Delta(0)|^2+b  |\Delta(0)|^4\\\nonumber
&- \Omega_{\mathrm{F}}^2\sum_{N}\Pi(0,N) [\Delta(0)\langle\Delta^*(N)\rangle+\langle\Delta(N)\rangle\Delta^*(0)].
\end{align}
Note that the functional integrals over $\Delta(N)$ for $N\neq 0$ in the numerator and denominator on the right hand side of Eq. (\ref{MFSC_equation}) cancel each other in the mean-field approximation.
Subsequently, under the condition of $\langle\Delta(N)\rangle \rightarrow 0$, we simplify Eq. (\ref{MFSC_equation}), expanding the expression $\exp\{-S_{\mathrm{MF}}[\Delta, \Delta^*]\}$ in powers of the long-rang coupling term to first order in $\Delta(0)$. We perform integration over $\Delta(0)$ and $\Delta^*(0)$ in the cylindrical coordinate representation; as a result, Eq. (\ref{MFSC_equation}) yields
\begin{align}\label{SC_equation}
1 =  \frac{\langle |\Delta(0)|^2\rangle}{T} \Omega_{\mathrm{F}}^2\sum_N \Pi(0,N),
\end{align}
where
\begin{eqnarray}\label{average}
\langle |\Delta(0)|^2\rangle =\frac{\int_0^{\infty} dx x e^{-\frac{1}{T} [(\frac{1}{U}-\frac{1}{U_c})\frac{\Omega_{\mathrm{F}}}{\Omega_a}x + bx^2] } }{\int_0^{\infty} dx e^{-\frac{1}{T} [(\frac{1}{U}-\frac{1}{U_c})\frac{\Omega_{\mathrm{F}}}{\Omega_a}x + bx^2] } }.
\end{eqnarray}
The solution of Eq. (\ref{SC_equation}) gives us a superconducting transition temperature $T_c$.
To investigate the solution of Eq. (\ref{SC_equation}) in the regimes of weak and strong interactions, we shall further consider the continuum limit by substituting
$\Omega_{\mathrm{F}} \sum_N \rightarrow \int d\mathbf{R}$.

\subsection{Weak-coupling regime}
In the weak-coupling limit $U \ll U_c$, we can neglect the temperature dependence in $\mathcal{U}_c(T)$ and set $\mathcal{U}_c(T)\equiv U_c$. Thus, at $U^{-1}-U_c^{-1}\gg  \frac{\Omega_a}{\Omega_{\mathrm{F}}} \sqrt{b T}\propto \sqrt{T/U_c^3}$ in the case when the system is far from the local pair binding, we can also neglect the $b$ term in (\ref{average}) and obtain $\langle |\Delta(0)|^2\rangle = T \frac{UU_c}{U_c-U} \frac{\Omega_a}{\Omega_{\mathrm{F}}}$. 
Consequently, we reproduce the standard BCS result (\ref{Tc_naive}) for the superconductor-metal transition temperature.

%%%%%%%%%%%%%%%%%%%%%% BEGIN FIGURE %%%%%%%%%%%%%%%%%%%
\begin{figure}[t]
\centering
\includegraphics[width=7cm]{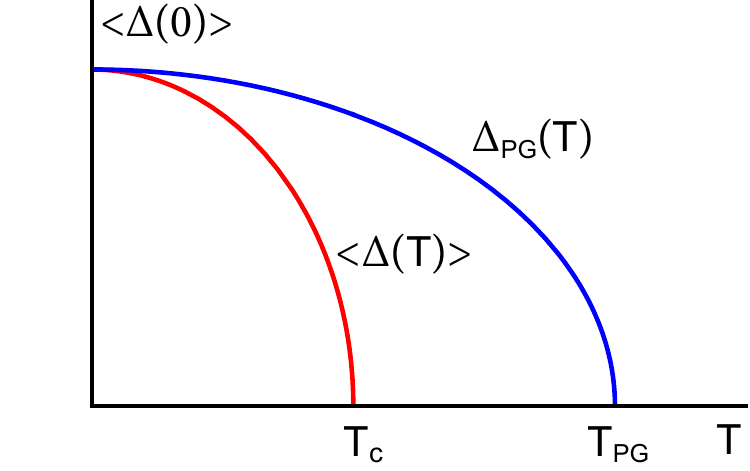}
\caption{\label{fig2} Schematics of the phase transition between a zero-temperature superconducting state and a high-temperature normal metal state via a preformed-pair state.}
\end{figure}
%%%%%%%%%%%%%%%%%%%%% END FIGURE %%%%%%%%%%%%%%%%%%%

\subsection{Strong-coupling regime}
In the strong-coupling regime, in the vicinity of the transition $\mathcal{U}^{-1}_c(T)-U^{-1}> \frac{\Omega_a}{\Omega_{\mathrm{F}}} \sqrt{b T} $, we now obtain
$\langle |\Delta(0)|^2\rangle = \frac{1}{2b} \frac{\Omega_{\mathrm{F}}}{\Omega_{a}}(\mathcal{U}_c^{-1}(T)-U^{-1}) \equiv |\Delta_{\mathrm{PG}}|^2$. 
 As a result, the transition temperature between the preformed Cooper pair state and the superconducting state yields
\begin{equation}\label{Very_main_results}
T_c= 2 |\Delta_{\mathrm{PG}}|^2\Omega_{\mathrm{F}} \nu\ln|\mu/\pi^2 T_c|.
\end{equation}
When $(\lambda_{F}/a)(1-U_{c}/U)\geq 1$ in (\ref{Very_main_results}), the prefactor before the logarithm becomes large, and the solution of (\ref{Very_main_results}) is given by $T_{c}\simeq \mu/\pi^{2}$. It qualitatively coincides with
(\ref{Tc_naive}) in the limit $U\rightarrow U_c$. The schematic phase diagram is presented in Fig. (\ref{fig1}).

%%%%%%%%%%%%%%%
\subsection{Low-temperature limit at strong coupling}
%%%%%%%%%%%%%%%
The theory of superconductivity at strong coupling is developed for temperatures much larger than the superconducting gap. 
However, we note that at large distances, $\Pi(\mathbf{R})$, which describes Andreev coupling among the preformed pairs, is cut by the thermal length $R\sim \ell_{\mathrm{T}}$ and diverges at $T\rightarrow 0$. Within the textbook BCS model, at zero temperature, the cutoff changes as $T\rightarrow |\Delta|$. Similarly, under the substitution $T\rightarrow \sqrt{T^{2}+|\Delta_{\mathrm{PG}}|^2}$, our approach might be continued to zero temperature as well.

In this low-temperatures regime, the superconducting energy gap is given by expression (\ref{average}), leading to $\langle \Delta \rangle = \Delta_{\mathrm{PG}}$. Noting different temperature dependences of $\langle \Delta \rangle$ and $\Delta_{\mathrm{PG}}$, we find an interesting phase diagram, as shown in Fig. (\ref{fig2}). Namely, the averaged order parameter vanishes at the transition to the preformed pair state $T \rightarrow T_c$ as
\begin{equation}
\langle \Delta \rangle \sim \sqrt{1-T/T_c},
\end{equation}
while $\Delta_{\mathrm{PG}}$ decreases only at $T\sim T_{\mathrm{PG}}> T_c$. This is a two-stage process: With increasing temperature, the phase disorder terminates the superconducting state. As the temperature increases further, the number of prebound electron states decreases, terminating local $\Delta(\textbf{R})$. Let us also emphasize that we keep the quartic term $b$ not because of the large temperatures, but because of the small parameter $|U-U_{c}|\ll U_{c}$. 

We also note that our calculations provide the upper bound for the critical temperature and an investigation of quantum and thermal fluctuations at strong interaction is beyond the present work.

%%%%%%%%%%%%%%%
\subsection{Fixed chemical potential or particle density} 
%%%%%%%%%%%%%%%
Above, we considered the case of a fixed chemical potential $\mu $. However, if we are interested in maintaining a constant particle number, we must account for the shift in the chemical potential at a fixed particle density.
For $T>T_{c}$, the local part of (\ref{Mean_field_functional}) determines the correction to the thermodynamical potential
as
\begin{equation}
\frac{\delta\Omega_{0}}{V}=-\frac{\Omega_{\mathrm{F}}}{4b} \left\{\Omega_{a}^{-1}[U^{-1} - \mathcal{U}^{-1}_{c}(T)]\right\}^{2}.
\end{equation}
The correction to the particle number  due to the attraction between electrons is given by 
\begin{equation}
\frac{\delta N}{V} \sim \frac{1}{\Omega_{\mathrm{F}}} \left[\frac{\lambda_{\mathrm{F}}}{a}\left(1 - \frac{U_{c}}{U}\right)\right]^{2}.
\end{equation}
For $(\lambda_\mathrm{F}/a)(1-U_{c}/U)\geq 1$, this correction can contribute a significant part of the particle density, leading to a reduction in $\mu$.
This was discussed in detail in \cite{Leggett, JLTP_Schmitt}.

%%%%%%%%%%%%%%%%
\section{\label{sec:pseudogap}Pseudogap phase}
%%%%%%%%%%%%%%%%%

Having identified the phase-incoherent preformed pair state within the temperature range $T_{\mathrm{PG}}>T>T_c$ under the strong-coupling conditions ($U>U_c$), we shall now investigate the fundamental spectral properties of quasiparticles in this intriguing state.

%%%%%%%%%%%%%%%%%%%%%% BEGIN FIGURE %%%%%%%%%%%%%%%%%%%
\begin{figure}[t]
\centering
\includegraphics[width=6cm]{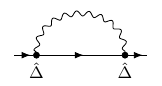}
\caption{\label{fig3} The self-energy due to electron scattering by random phase disorder within the Born approximation in the Nambu representation and in the static limit (no frequency transfer via the bosonic correlator). 
The vertex represents a matrix in Nambu space $\hat{\Delta} = \mathrm{Re}[\Delta] \tau_x - \mathrm{Im}[\Delta]\tau_y$, where $\tau_{x,y,z}$ are the Pauli matrices in Nambu space and $\Delta =  \mathrm{Re}[\Delta] + i  \mathrm{Im}[\Delta]$, while the solid line stands for the Green function in Nambu space, $\hat{G}_{\omega_n}(\mathbf{p}) = G_{\omega_n}(\mathbf{p})(1+\tau_z)/2 - G_{-\omega_n}(\mathbf{p})(1-\tau_z)/2$.}
\end{figure}
%%%%%%%%%%%%%%%%%%%%% END FIGURE %%%%%%%%%%%%%%%%%%%

As we discussed above, the preformed Cooper pair state can be seen as a cluster of domains with a spatially random phase of the order parameter $\Delta(\mathbf{R}_N)$.
The back-action of the random phase on electrons can be studied in analogy to the problem of electron scattering by random disorder in metals. 
In our case the scattering of quasiparticles on $\Delta(\mathbf{R}_N)$ with a vanishing average $\langle \Delta \rangle =0$ due to random phase is described by the Hamiltonian
$
H_{\mathrm{int}} =\Omega_{\mathrm{F}} \sum_N[ \Delta(\mathbf{R}_N)\psi^{\dag}_{\uparrow}(\mathbf{R}_N) \psi^{\dag}_{\downarrow}(\mathbf{R}_N) + \mathrm{h.c.}]
$.

At $T\rightarrow T_c$, to evaluate the correlation function of the scattering potential, we express $\Delta(\mathbf{R}_N) = |\Delta_{\mathrm{PG}}| e^{i\phi(\mathbf{R}_N)}$ and, similar to Eq. (\ref{MFSC_equation}), obtain
\begin{eqnarray}\label{Disorder_correlator}
\frac{\langle\Delta(0) \Delta^*(\mathbf{R})\rangle}{|\Delta_{\mathrm{PG}}|^2}
=  \frac{\int_0^{2\pi} d\phi \cos(\phi) \exp\{ E_{\mathrm{J}}(R)  \cos(\phi )/T\}  }{\int_0^{2\pi} d\phi \exp\{ E_{\mathrm{J}}(R)  \cos(\phi )/T\} },~~
\end{eqnarray}
where $E_{\mathrm{J}}(R)$ is the energy of Josephson coupling between the two domains 
\begin{equation}
E_{\mathrm{J}}(R) = T \frac{2\pi \mathcal{L}_{\mathrm{T}}^3}{\ell_{\mathrm{T}}R^2}\frac{1}{\sinh(2\pi R/\ell_{\mathrm{T}})}.
\end{equation}
The length $\mathcal{L}_{\mathrm{T}} = \lambda_{\mathrm{F}}(2|\Delta_{\mathrm{PG}}|^2/T\mu)^{1/3}$ is much smaller than the thermal correlation length $\ell_{\mathrm{T}}$,  $\mathcal{L}_{\mathrm{T}}/\ell_{\mathrm{T}}\propto (|\Delta_{\mathrm{PG}}| T/\mu^2)^{2/3}\ll1$.

The correlation function (\ref{Disorder_correlator}) can be written as the ratio of two modified Bessel functions of the first kind:
\begin{equation}\label{correlator_real}
\langle\Delta(0) \Delta^*(\mathbf{R})\rangle = |\Delta_{\mathrm{PG}}|^2 \frac{I_1(E_{\mathrm{J}}(R)/T)}{I_0(E_{\mathrm{J}}(R)/T)}.
\end{equation} 
Using the asymptote of the Bessel functions, we observe that the phase exhibits slow changes at small length scales $R < \mathcal{L}_{\mathrm{T}}$, namely, $\langle e^{i(\phi_{0}-\phi_{N}}\rangle \simeq 1$. 
However, in the case $\ell_{\mathrm{T}}>R > \mathcal{L}_{\mathrm{T}}$, the correlation function decays as a power law $\propto \mathcal{L}_{\mathrm{T}}^3/R^{3}$, 
while at $R>\ell_{\mathrm{T}}$ the decay is exponential, $\propto \mathcal{L}_{\mathrm{T}}^3 \exp(-2\pi R/\ell_{\mathrm{T}})/(\ell_{\mathrm{T}}R^{2})$, indicating a phase loosening at larger distances.

We are now in the position to evaluate the correction to the density of states of quasiparticles due to scattering by phase disorder, given by the diagram in Fig. (\ref{fig3}).
Performing Fourier transformation of the correlation function $\langle\Delta\Delta^*\rangle_{\mathbf{q}} = \int d\mathbf{R} \langle\Delta(\mathbf{R}) \Delta^*(0)\rangle e^{-i\mathbf{q}\cdot\mathbf{R}} $ in (\ref{correlator_real}), we obtain an equation for the Green function in the Born approximation $[i\omega_n - \xi_{\mathbf{p}} - \Sigma_{\omega_n}(\mathbf{p})]^{-1} G_{\omega_n}(\mathbf{p})=1$, where the self-energy is given by
\begin{eqnarray}
\Sigma_{\omega_n}(\mathbf{p}) = \int \frac{d^3q}{(2\pi)^3} \frac{ \langle\Delta \Delta^*\rangle_{\mathbf{p}-\mathbf{q}}}{i\omega_n+\xi_{\mathbf{q}}}.
\end{eqnarray}
Noting that $\mathcal{L}_{\mathrm{T}}/\lambda_{\mathrm{F}}< 1$, or, equivalently, $|\Delta_{\mathrm{PG}}| < \sqrt{T\mu}$, we can take the asymptote of  the correlation function $\langle\Delta(0) \Delta^*(\mathbf{R})\rangle \sim R^{-3}$.
The correction to the electron density of states per spin due to electron scattering by phase disorder is given by
\begin{equation}
\delta\nu (\omega) = -\frac{1}{\pi}\mathrm{Im}\int\frac{d^3p}{(2\pi)^3}\frac{ \Sigma_{\omega+i\delta}(\mathbf{p})}{(\omega - \xi_{\mathbf{p}}+i\delta)^2},
\end{equation}
which can be brought to the form
\begin{equation}
\frac{\delta\nu (\omega)}{\nu} =  -\frac{4|\Delta_{\mathrm{PG}}|^2}{v^2_{\mathrm{F}}\sqrt{1+\omega/\mu}}\int_0^{\infty} \frac{I_1(\frac{E_{\mathrm{J}}(r)}{T})}{I_0(\frac{E_{\mathrm{J}}(r)}{T})}\cos\left(\frac{2\omega}{v_{\mathrm{F}}} r\right)r dr ,
\end{equation}
At frequencies $\mu>|\omega|$, subtracting the part, $\delta\nu (0)/\nu  = -4\pi^2 |\Delta_{\mathrm{PG}}|^4/(T\mu^3)$, we obtain
\begin{equation}\label{pseudogap_eq}
\frac{1}{\nu}[\delta\nu (\omega)- \delta\nu (0)] = 2\pi^4 \frac{|\Delta_{\mathrm{PG}}|^4}{\mu^4} \frac{|\omega|}{T}.
\end{equation}
Linear dependence emerges from the asymptotic behaviour of the correlation function, $\langle \Delta(0)\Delta^{*}({\bm R}) \rangle \sim R^{-3}$, at intermediate distances. This dependence becomes more singular when correlations decay more slowly. 
The frequency-independent correction has a negative sign, which together with (\ref{pseudogap_eq}) leads to a dip in the density of states, resembling the pseudogap behaviour.
It is worth noting that, in the case of a two-dimensional superconductor, the correction follows a linear frequency dependence as well $\delta\nu (\omega)- \delta\nu (0) \propto |\omega|/T$.
As the temperature approaches $T\rightarrow T_{\mathrm{PG}}$, the correlator (\ref{Disorder_correlator}) is no longer applicable. Instead, the disorder averaging technique within the Gaussian approximation results in the scattering time. In conclusion, we find a pseudogap state at temperatures above the superconducting transition temperature $T\rightarrow T_c$.

% ==============

\section{\label{sec:conclusion}Conclusion}
To summarize, we developed a model of superconductivity in conductors with a parabolic band spectrum and strong attractive interactions between electrons. 
We highlighted the significance of spatial scale separation in analyzing Cooper pair instability.

We demonstrated that a strong interaction potential, which can bind electrons into pairs, results in a state characterized by a spatially random phase of the order parameter. 
This inhomogeneous state lacks global phase coherence. However, it can be characterized by the pseudogap arising from electron scattering by phase disorder. At low temperatures, long-range correlations between regions with different phases plays an important role. These correlations drive phase synchronization, and therefore superconductivity, in the system. We emphasized that the preformed Cooper pair model in conductors with parabolic bands and strong interactions can be mapped onto the model of superconductivity in flat-band systems.

Finally, we note that at the transition $U \gg U_{c}$, in the case of a fixed particle number, the chemical potential approaches zero, $\mu\rightarrow 0$, causing the contribution of nonlocal terms to vanish. Therefore, we expect a different superconducting state compared to the one considered here. 

% ==============
\section{Acknowledgements} We would like to thank Mikhail Feigel'man for raising the question of the Ginzburg-Levanyuk criterion for the validity of the mean-field theory.
We thank the Pirinem School of Theoretical Physics, where our research was initiated, for warm hospitality.
% ==============

\appendix
\section{\label{sec:Appendix_A} Self-consistency equation}

The linearized self-consistency equation is given by
\begin{align}\nonumber
&\Delta(\mathbf{r},\mathbf{r}') \Theta(a-|\mathbf{r}-\mathbf{r}'|) = UT\sum_n\int d\mathbf{r}_1d\mathbf{r}_2 
\Theta(a-|\mathbf{r}_1-\mathbf{r}_2|)\\
&\times
\Theta(a-|\mathbf{r}-\mathbf{r}'|)G_{\omega_n}(\mathbf{r}-\mathbf{r}_1) \Delta(\mathbf{r}_1,\mathbf{r}_2) G_{-\omega_n}(\mathbf{r}'-\mathbf{r}_2),
\end{align}
where
\begin{align}
G_{\omega_n}(\mathbf{R}) = -\frac{m}{2\pi R} \exp\left\{i\mathrm{sign}(\omega_n)\sqrt{1+\frac{i\omega_n}{\mu}}\frac{2\pi R}{\lambda_{\mathrm{F}}}\right\}
\end{align}
is the Green function in the Matsubara representation and $\lambda_{\mathrm{F}}= 2\pi/\sqrt{2m\mu}$ is the Fermi wavelength. Transforming to the center-of-mass $\mathbf{R}$ and relative $\mathbf{r}$ coordinates, we obtain
\begin{align}\nonumber
&\Delta(\mathbf{R},\mathbf{r})= U T\sum_n\int d\mathbf{R}' \int_{r'<a}d\mathbf{r}' \Theta(a-r) \Delta\left(\mathbf{R}',\mathbf{r}'\right)
 \\
 &\times 
 G_{\omega_n}\left(\mathbf{R}-\mathbf{R}'+\frac{\mathbf{r}-\mathbf{r}'}{2}\right)  G_{-\omega_n}\left(\mathbf{R}-\mathbf{R}' - \frac{\mathbf{r}-\mathbf{r}'}{2}\right).
\end{align}
As in the main text, we introduce
\begin{eqnarray}
\Pi(\mathbf{R},\mathbf{r}) = T\sum_n G_{\omega_n} \left(\mathbf{R}+\mathbf{r}\right)G_{-\omega_n}\left(\mathbf{R}-\mathbf{r}\right).
\end{eqnarray}
Formal rewriting gives
\begin{align}\label{SM_mainequation}\nonumber
&\Delta(\mathbf{R},\mathbf{r})= U \int d\mathbf{R}' \int_{r'<a}d\mathbf{r}'
 \Theta(a-r)
 \\
 &\times \Pi\left(\mathbf{R}-\mathbf{R}', \frac{\mathbf{r}-\mathbf{r}'}{2}\right)  \Delta\left(\mathbf{R}',\mathbf{r}'\right).
\end{align}
We separate the local and nonlocal contributions. We start with the local term. At zero temperature, we find
\begin{align}\label{intermediate_SM}\nonumber
&\Pi(\mathbf{R},\mathbf{r}) = T\sum_n \frac{m^2}{4\pi^2}\frac{1}{|\mathbf{R}+\mathbf{r}||\mathbf{R}-\mathbf{r}|} \\\nonumber
&\times  \exp\left\{ i\mathrm{sign}(\omega_n)[\sqrt{2im\omega_n} |\mathbf{R}+\mathbf{r}| - \sqrt{-2im\omega_n} |\mathbf{R}-\mathbf{r}| ]\right\}
\\
&=\frac{m}{2\pi^3}\frac{1}{(|\mathbf{R}+\mathbf{r}|^2+|\mathbf{R}-\mathbf{r}|^2)^2}.
\end{align}
At $R<\lambda_{\mathrm{F}}$, the temperature-dependent correction can be evaluated as
\begin{eqnarray}\nonumber
\delta\Pi(\mathbf{R},\mathbf{r}) &=& \frac{m^2}{4\pi^2R^2}\int_{-\mu}^{\mu}\frac{d\epsilon}{2\pi}\left[\tanh\left(\frac{\epsilon}{2T}\right)-\mathrm{sign}(\epsilon)\right]
\\\nonumber
&\times& \sin\{p_{\mathrm{F}}R[\sqrt{1+\epsilon/\mu}-\sqrt{1-\epsilon/\mu} ]\}
\\
& \sim& -\frac{m^{3}\lambda_{F}T^{2}}{24\pi^{2}R}.
\end{eqnarray}

The nonlocal contribution can be evaluated as
\begin{align}
&\Pi(\mathbf{R},\mathbf{r}) = T\sum_{n>0} \frac{m^2}{2\pi^2 |\mathbf{R}+\mathbf{r}| |\mathbf{R}-\mathbf{r}|}
\\\nonumber
&\times \cos\left\{\frac{2\pi}{\lambda_{\mathrm{F}}}(|\mathbf{R}+\mathbf{r}| - |\mathbf{R}-\mathbf{r}|)\right\} 
\\\nonumber
&\times
\exp\left\{-\frac{\omega_n}{v_{\mathrm{F}}}(|\mathbf{R}+\mathbf{r}| + |\mathbf{R}-\mathbf{r}|)\right\} 
\\\nonumber
&=  \frac{m^2 T}{4\pi^2 |\mathbf{R}+\mathbf{r}| |\mathbf{R}-\mathbf{r}|} \frac{\cos\left\{\frac{2\pi}{\lambda_{\mathrm{F}}}(|\mathbf{R}+\mathbf{r}| - |\mathbf{R}-\mathbf{r}|)\right\} }{\mathrm{sinh}(\pi (|\mathbf{R}+\mathbf{r}| + |\mathbf{R}-\mathbf{r}|)T/v_{\mathrm{F}})}.
\end{align}
At small distances $|\mathbf{R}\pm \mathbf{r}|\rightarrow \lambda_{\mathrm{F}}$, the spatial dependence in $\mathrm{sinh}(\pi (|\mathbf{R}+\mathbf{r}| + |\mathbf{R}-\mathbf{r}|)T/v_{\mathrm{F}}) \rightarrow \mathrm{sinh}(2\pi \lambda_{\mathrm{F}}T/v_{\mathrm{F}})$ is cut by the Fermi wave-length. Hence, we can substitute $|\mathbf{R}\pm \mathbf{r}| = R$ in this term in what follows.

\subsection{Summary}
At $\lambda_{F}>R$, the kernel is determined by the local term
\begin{equation}
\Pi(\mathbf{R},\mathbf{r}) = 
\frac{m}{2\pi^3}\frac{1}{(|\mathbf{R}+\mathbf{r}|^2+|\mathbf{R}-\mathbf{r}|^2)^2}-\frac{m^{3}\lambda_{F}T^{2}}{24\pi^{2}R}
\end{equation}

 At larger distances $R> \lambda_{F}$, the kernel is determined by the nonlocal term
\begin{equation}
\Pi(\mathbf{R},\mathbf{r})=\frac{m^2 T}{4\pi^2 R^{2}} \frac{1} {\mathrm{sinh}(2\pi RT/v_{\mathrm{F}})}.
\end{equation}

There are two different regimes which will be justified a posteriori, namely, the weak coupling $U\ll \pi^2/4ma^2$ and strong coupling $U> \pi^2/4ma^2$. 
At $U\ll\pi^2/4ma^2$, we have a standard BCS regime. The nonlocal term contributes to the self-consistency equation and Cooper instability. The nonlocal term can be considered as perturbation.
On the other hand, at $U> \pi^2/4ma^2$, as we will see below, the solution of the self-consistency equation is determined by the local contribution. The nonlocal term can be treated as a perturbation. In the situation when the local contribution
determines the Cooper instability, it suffices to discretize the system into small regions, each with volume $\lambda_\mathrm{F}\times \lambda_\mathrm{F}\times \lambda_\mathrm{F}$. Neglecting the nonlocal contribution means that the self-consistency equation for the Cooper instability can be considered independently in each region. Since the order parameter is determined up to a phase, it has an uncorrelated phase in different regions.

\subsection{Strong coupling}
At strong coupling the nonlocal contribution can be considered as perturbation. 
We can focus on only a single region. 
We seek for a solution $\psi(\mathbf{r})$ of Eq. (\ref{Main_equation}) in the s-wave channel in the form:
\begin{equation}\label{Expansion_SM}
\Delta(\mathbf{R},\mathbf{r}) = \Delta(\mathbf{R}) \psi(\mathbf{r}).
\end{equation}
Considering that $\Delta(\mathbf{R})$ changes slowly on the Fermi wave length, we set $\Delta(\mathbf{R}) \approx \Delta(\mathbf{R}')$.
We note that $|\mathbf{R}-\mathbf{R}'| $ is bounded by the typical size of the preformed Cooper pair region given by the Fermi wave-length. Hence we integrate the local term as
\begin{equation}
\int_{R<\lambda_{\mathrm{F}}}
\frac{d\mathbf{R}}{(R^2+|\mathbf{r}-\mathbf{r}'|^2/4)^2} = \frac{2\pi^2}{|\mathbf{r}-\mathbf{r}'|}.
\end{equation}
Here, we neglect term $\propto \lambda^{-1}_{\mathrm{F}}$, which might be summed up into the long-distance part. As a result, the self-consistency equation can be brought to the form

\begin{eqnarray}\nonumber
\psi(\mathbf{r}) &=& \frac{\pi\Theta(a-r)}{16} \frac{U}{U_c} \\
&\times&
\int_{r'<a}d\mathbf{r}'
 \left(\frac{1}{|\mathbf{r}-\mathbf{r}'|} -  \frac{1}{6}  m^{2}\lambda^{3}_{\mathrm{F}}T^{2}  \right)\psi(\mathbf{r}'),~~~~
\end{eqnarray}
where 
\begin{equation}
U_c = \frac{\pi^2}{4ma^2}
\end{equation}
is the critical interaction strength. After integrating  over the angle
$\psi(r) = \int \frac{d\vec{\theta}}{4\pi} \psi(\mathbf{r})$, we obtain
\begin{eqnarray}\label{S}\nonumber
\psi(r) &=& \frac{\pi^2}{4a^{2}} \frac{U}{U_c} \int_{0}^{a}dr' r'^2 
\\
&\times&\left(\frac{r+r'-|r-r'|}{2rr'} -  \frac{ 1}{6} m^{2}\lambda^{3}_{\mathrm{F}}T^{2}  \right)\psi(r').~~~~~
\end{eqnarray}

From (\ref{S}), for $r\psi(r)$, we have the differential equation
\begin{equation}
\frac{d^{2}}{dr^{2}}[r\psi(r)]=-\frac{\pi^{2}U}{4a^{2}U_{c}}[r\psi(r)]
\end{equation}
with the boundary condition for the first derivative at $r=a$:

\begin{eqnarray}\nonumber
\frac{d}{dr}(r\psi)|_{r=a} =&-&\frac{\pi m^{2}\lambda^{3}_{\mathrm{F}}T^{2} U}{24U_{c}a^{2}}\int^{a}_{0}dr'r'^{2}\psi(r').
\end{eqnarray}
A nonzero solution is
\begin{equation}
\psi(\mathbf{r}) = \frac{1}{r} \sin\left(\frac{\pi r}{2a}\sqrt{\frac{U}{U_c}}\right)
\end{equation}
at
\begin{equation}
\frac{U}{U_c} = 1+ \frac{(4\pi)^{2}a}{3\lambda_\mathrm{F}}\left(\frac{T}{\mu}\right)^{2}.
\end{equation}

As a result, at $U\gtrsim U_c$, we obtain the crossover temperature for the formation of the local Cooper pairing,
\begin{equation}
T_{\mathrm{PG}} = \frac{\mu}{4\pi}\left(\frac{3\lambda_{\mathrm{F}}}{ a}\right)^{1/2} \left(\frac{U}{U_c}-1\right)^{1/2}.
\end{equation}

\subsection{Weak-coupling limit}
In the limit $U\ll U_c$, we obtain
\begin{align}
&\int_{r<a}d^3r \Delta(\mathbf{R},\mathbf{r}) 
=
\\\nonumber
&=U \int_{r,r'<a}d^3r d^3r' \left\{\frac{m}{4\pi|\mathbf{r}-\mathbf{r}'|} -\nu \ln \left|\frac{\pi\lambda_{\mathrm{F}}}{\ell_{\mathrm{T}}} \right| \right\}\Delta(\mathbf{R},\mathbf{r}').
\end{align}
Using 
\begin{equation}
\Delta(\mathbf{R},\mathbf{r}) = \frac{ \Delta(\mathbf{R})}{r} \sin\left(\frac{\pi r}{2a}\sqrt{\frac{U}{U_c}}\right),
\end{equation}
we obtain the equation
\begin{equation}
1- \frac{U}{U_c}= -U\nu \Omega_a \ln \left|\frac{\pi\lambda_{\mathrm{F}}}{\ell_{\mathrm{T}}} \right|,
\end{equation}
where $\Omega_a= 4\pi a^3/3$. The above equation gives the superconducting transition temperature
\begin{equation}\label{Sup_transition_temperature_ours}
T_c = \frac{\mu}{\pi^2}\exp\left\{-\frac{1-U/U_c}{U\nu \Omega_a}\right\},
\end{equation}
which is expression (\ref{Tc_naive}) in the main text. To compare Eq. (\ref{Sup_transition_temperature_ours}) with Gor'kov and Melik-Barkhudarov renormalization \cite{Barkhudarov_Tc}, we note that
\begin{equation}
\left(\frac{1}{U} - \frac{1}{U_c} \right) \frac{1}{\Omega_a^3}  \equiv \frac{1}{|f_0|},
\end{equation}
where $f_0$ is the scattering amplitude, which is calculated in the second-order scattering approximation, as done in Ref. \cite{Barkhudarov_Tc} and reviewed in Ref. \cite{Ohashi_progress}. 

%%%%%%%%%%%%%%%%%%%%%%%%%%%
\bibliography{Paraband_references_PRB.bib}
%%%%%%%%%%%%%%%%%%%%%%%%%%%

\end{document}